\documentclass[12pt]{article}
\usepackage{latexsym}
\usepackage{amssymb}
\usepackage{amsmath}
\usepackage{graphicx}
\usepackage{enumitem}
\usepackage{slashed}
\usepackage{hyperref}

\def\be{\begin{equation}}
\def\ee{\end{equation}}
\def\bear{\begin{eqnarray}}
\def\eear{\end{eqnarray}}
\def\nn{\nonumber}

\newcommand\bra[1]{{\langle {#1}|}}
\newcommand\ket[1]{{|{#1}\rangle}}

\def\dd{\mbox{d}}

\def\bra{\langle}
\def\ket{\rangle}

\newcommand{\ti}[1]{\tilde{#1}}

\newcommand{\sm}[1]{\mbox{\scriptsize #1}}
\newcommand{\tn}[1]{\mbox{\tiny #1}}
\renewcommand{\@}[1]{\sqrt{#1}}
\renewcommand{\le}[1]{\label{#1}\end{eqnarray}}
\newcommand{\bea}{\begin{eqnarray}}
\newcommand{\eea}{\end{eqnarray}}

\newcommand{\eq}[1]{(\ref{#1})}
\def\nn{\nonumber\\}

\def\ffract#1#2{\raise .35 em\hbox{$\scriptstyle#1$}\kern-.25em/
\kern-.2em\lower .22 em \hbox{$\scriptstyle#2$}}

\begin{document}

\pagestyle{empty}

\centerline{{\Large \bf The Empirical Under-determination Argument}}
\vskip.7cm
\centerline{{\Large \bf  Against Scientific Realism for Dual Theories}}
\vskip.9cm

\begin{center}
{\large Sebastian De Haro}\\
\vskip .7truecm
{\it Institute for Logic, Language of Computation, University of Amsterdam}\footnote{Forthcoming in {\it Erkenntnis.}}\\
{\it Institute of Physics, University of Amsterdam}\\
{\it Vossius Center for History of Humanities and Sciences, University of Amsterdam}

\vskip .7truecm
{\tt s.deharo@uva.nl}
\vskip 2cm
\today
\end{center}

\vskip 7truecm

\begin{center}
\textbf{\large \bf Abstract}
\end{center}

This paper explores the options available to the anti-realist to defend a Quinean empirical under-determination thesis using examples of dualities. I first explicate a version of the empirical under-determination thesis that can be brought to bear on theories of contemporary physics. Then I identify a class of examples of dualities that lead to empirical under-determination. But I argue that the resulting under-determination is benign, and is not a threat to a cautious scientific realism. Thus dualities are not new ammunition for the anti-realist. The paper also shows how the number of possible interpretative options about dualities that have been considered in the literature can be reduced, and suggests a general approach to scientific realism that one may take dualities to favour.

\newpage
\pagestyle{plain}

\tableofcontents

\newpage

\section{Introduction}\label{intro}

Over the last twenty to thirty years, dualities have been central tools in theory construction in many areas of physics: from statistical mechanics to quantum field theory to quantum gravity. A duality is, roughly speaking, a symmetry between two (possibly very different-looking) theories. So in physics: while a symmetry typically maps a state of the system into another appropriately related state (and likewise for quantities); in a duality, an entire theory is mapped into another appropriately related theory. And like for symmetries, there is a question of under what conditions dual theories represent empirically equivalent situations. Indeed, in some cases, physicists claim that dual pairs of theories describe the very same {\it physical,} not just the very same {\it empirical,} facts (more on this in Section \ref{secdual}).

Thus there is a natural question of {\it whether dualities can generate interesting examples of empirical under-determination,} or under-determination of theory by empirical data.

The empirical under-determination thesis says that, roughly speaking, `physical theory is underdetermined even by all {\it possible} observations... Physical theories... can be logically incompatible and empirically equivalent' (Quine, 1970:~p.~179).\footnote{The qualification `all possible' distinguishes this type of under-determination thesis from the under-determination of theories by the available evidence {\it so far,} also called `transient under-determination' (Sklar, 1975:~p.~380), which is of course a very common phenomenon. Transient under-determination tends to blur the distinction between the limits of our current state of knowledge and understanding of a theory vs.~the theory's intrinsic limitations. Furthermore, under-determination by {\it all possible} evidence is closer to discussions of dualities. For these reasons, in this paper I restrict attention to the original Quinean under-determination thesis. \label{sofar}}

Despite the wide philosophical interest of the empirical under-determination thesis, it is controversial whether there are any genuine examples of it (setting aside under-determination `by the evidence so far'). Quine himself regarded this as an `open question'. One well-known example is the various versions of non-relativistic quantum mechanics, including its different interpretations; however, it is controversial whether these are cases of under-determination by {\it all the possible} evidence, or by all the evidence {\it so far}. Likewise, Laudan and Leplin (1991:~p.~459) say that most examples of under-determination are contrived and limited in scope: 
\begin{quote}\small
It is noteworthy that contrived examples alleging empirical equivalence always involve the relativity of motion; it is the impossibility of distinguishing apparent from absolute motion to which they owe their plausibility. This is also the problem in the pre-eminent historical examples, the competition between Ptolemy and Copernicus, which created the idea of empirical equivalence in the first place, and that between Einstein and H.~A.~Lorentz.
\end{quote} 

But dualities certainly go well beyond relative motion and the interpretation of non-relativistic quantum mechanics. For they are at the centre of theory construction in theoretical physics: and so, if they turned out to give cases of under-determination, this would show that the problem of under-determination is at the heart of current scientific research. And since philosophers have thought extensively about under-determination, this old philosophical discussion could contribute to the understanding of current developments in theoretical physics.

Furthermore, empirical under-determination is also one of the arguments that have been mounted against scientific realism. Thus if dualities turn out to {\it not} illustrate under-determination, then this particular argument against scientific realism would be undermined. Either way, the question of whether dualities give cases of under-determination deserves scrutiny. 

So far as I know, and apart from an occasional mention, the import of dualities for the under-determination debate has so far only been studied in a handful of previous papers: see early papers by Dawid (2006, 2017a), Rickles (2011, 2017), Matsubara (2013) and recent discussions by Read (2016) and Le Bihan and Read (2018).\footnote{For a discussion of the import of dualities for the question of scientific realism, see Dawid (2017a).} Given that a rich Schema for dualities now exists equipped with the necessary notions for analysing under-determination,\footnote{See De Haro and Butterfield (2017), De Haro (2019, 2020), and Butterfield (2020).} and several cases of rigorously proven dualities also exist, the time is ripe for a study of the question of empirical under-determination vis-\`a-vis dualities.

In this paper, I aim to undertake a study of the main options available to produce examples of empirical under-determination using dualities. Thus, rather than giving a response to the various under-determination theses or defending scientific realism, my aim is to analyse whether dualities offer genuine examples of under-determination---as it is reasonable to expect that they do. We will find that, although the examples of dualities in physics {\it do} illustrate the under-determination thesis, they do so in a benign way, i.e.~they are not a threat to---cautious forms of---scientific realism. In particular, I will argue that---leaving aside cases of {\it transient} under-determination\footnote{Transient under-determination for dualities is discussed in Dawid (2006).}---not all of the interpretative options that have been considered by Matsubara (2013) and Le Bihan and Read (2018) are distinct or relevant options for the problem of {\it emprical} under-determination, and that ultimately there is a single justified option for the cautious scientific realist. Thus dualities may be taken to favour a cautious approach to scientific realism.

In Section \ref{eudt}, I introduce the Quinean empirical under-determination thesis. Section \ref{secdual} introduces the Schema for duality. Section \ref{DandU} then explores whether dualities give cases of under-determination, and whether scientific realism is in trouble. Section \ref{conclusion} concludes. %draws some general conclusions for the question of theory individuation in connection with scientific realism.

\section{The Empirical Under-determination Thesis}\label{eudt}

This Section introduces the Quinean empirical under-determination thesis, and the allied concepts of empirical and theoretical equivalence.

\subsection{Quine's under-determination thesis, and strategies for response}\label{Qbeudt}

The word `under-determination' is notoriously over-used in philosophy of science. The under-determination thesis that I am concerned with here is what Quine (1975:~p.~314) called `empirical under-determination', which is not to be confused with the Duhem-Quine thesis, or holism: indeed a quite different doctrine.\footnote{Lyre (2011:~p.~236) gives an interesting discussion of the relation and contrast between under-determination, on the one hand, and Duhem-Quine holism and Humean under-determination, on the other.\label{otherUD}} Quine gives several different formulations of his own thesis, of which the simplest version is:\\
\\
{\bf Quinean under-determination$_0$:} two theory formulations are under-determined if they are empirically equivalent but logically incompatible.\\ %\footnote{For short, I will often simply refer to this as `under-determination$_0$'.}\\
\\
Under-determination theses are prominent as arguments against scientific realism. For if the same empirical facts can be described by two theories that contradict one another, why should we believe what one of the theories says?\footnote{I concur with Stanford (2006:~p.~17) in rejecting the use of fictitious examples to put forward under-determination theses: `[T]he critics of underdetermination have been well within their rights to demand that serious, nonskeptical, and genuinely distinct empirical equivalents to a theory actually be produced before they withhold belief in it and refusing to presume that such equivalents exist when none can be identified.' \label{stanford}} 

There are, roughly, two strategies for realists to try to respond to the challenge of empirical under-determination, as follows:\\
\\
(i)~~One can try to undercut the under-determination threat through appropriate {\it modifications of one's notions of equivalence,} so that on the correct notions there is no under-determination after all, i.e.~what one thought were examples of empirical under-determination turn out not to be. Since under-determination$_0$ involves two notions of equivalence in tension (under-determination ``lives in the space between empirical and logical equivalence''), this can be done in two ways. First, one can try to argue that an appropriate notion of {\it empirical equivalence} is sufficiently strong, so that the class of potential examples of under-determination is reduced (i.e.~because the two theory formulations in the putative example are {\it not} empirically equivalent). Or one can try to argue that the appropriate notion of {\it logical equivalence} (or, more generally, theoretical equivalence: see Section \ref{Peq}) is sufficiently weak, so that the class of potential examples of under-determination is reduced (i.e.~because the criterion of logical equivalence adopted is liberal, and makes the pairs of theory formulations in the putative examples logically equivalent). \\
\\
(ii)~~One can try to {\it use alternative assessment criteria} that do not involve modifications of the notions of equivalence. These criteria may be empirical, super-empirical, or even non-empirical. They are often forms of ampliative inference which offer more ``empirical support'' to one theory than the other; or one theory may be ``better confirmed by the evidence than the other'' (Laudan, 1990:~p.~271). \\
\\
I take Quine's challenge of under-determination$_0$ to be primarily about (i) not (ii). This follows from the formulation of the challenge itself. Thus my position is that in so far as Quinean under-determination$_0$ poses a challenge to scientific realism, one must aspire to meet it in its own terms, i.e.~by adopting the first strategy. 

If (i) turns out to fail, so that we need to adopt strategy (ii) instead, this might still not be a blow to scientific realism, since (ii) can indeed solve the empirical under-determination problems that the practicing scientist might be concerned with. In particular, if we adopt the strategy (i), we may still be left with some cases of under-determination, and {\it then} we may, and should, still ask: which of these two theory formulations is better confirmed, and which one should we further develop or accept? That is, the strategy (ii) remains relevant.

Also, one should note that, although the strategy (i) is semantic and (ii) is epistemic, epistemic considerations are relevant to both approaches, since we are interested in theories that are proposed as candidate descriptions of the actual world. Thus in various parts of the paper, when I talk about holding a certain thesis (e.g.~giving a verdict of theoretical equivalence), I will also consider the justification that we have for this thesis.

Fortunately so far, for realists, cases of under-determination seem hard to come by. Quine (1975:~p.~327) could find no good examples, and concluded that it is an `open question' whether such alternatives to our best `system of the world' exist. (See also Earman (1993:~pp.~30-31)).\footnote{Laudan and Leplin's (1991:~p.~458) example, the ``bason'', illustrates how hard-pressed the group of philosophers involved in the empirical under-determination debate have sometimes been to find what remotely look like genuine physics examples. The ``bason'' is a mythical particle invented to detect absolute motion: it fortuitously arises as a result of absolute motion, and `the positive absolute velocity of the universe represents energy available for bason creation'. Unfortunately, this cunning philosophical thought experiment hardly deserves the name `scientific theory'. See my endorsement of Stanford's critique of the use of fictitious theories, in footnote \ref{stanford}.\label{bason}} As late as 2011, Lyre (2011:~p.~236) could write that empirical under-determination `suffers from a severe but rarely considered problem of missing examples'.\footnote{Lyre (2011:~pp.~237-241) contains a useful classification of some available examples. I agree with Lyre's verdict about most of these examples being cases of transient under-determination, so that `in retrospect such historic cases appear mainly as artefacts of incomplete scientific knowledge---and do as such not provide really worrisome... cases' of empirical under-determination (p.~240).\label{lyre}}\\ 

In the rest of the paper, I will discuss strategy (i), and how dualities address this problem of missing examples. As we will see, my analysis will in effect be as follows. (A)~I will only minimally strengthen the notion of empirical equivalence (in Section \ref{emeq}), although I will also use this notion in a liberal way (in Section \ref{eed}). (B)~I will also {\it strengthen} the notion of equivalence that is to replace Quine's condition of logical equivalence. Thus the effects of my two modifications on the empirical under-determination thesis are in tension with each other. (A) strengthens empirical equivalence, which should help strategy (i) (cf.~the beginning of this Section), but I will also allow a liberal use of it, which again runs against (i). (B) is a substantial strengthening of the logical criteria of equivalence. So the overall effect of my analysis will not necessarily be good news for scientific realists---and this is of course the case we should consider, if we are to make the challenge to scientific realism as strong as possible. My liberal use of (A) will be motivated by how physicists use dualities, and by a discussion of the semantic and syntactic conceptions of empirical equivalence. The strengthening (B) will also be based on work on dualities, and how dualities give us a plausible criterion of theory individuation, that was developed independently of the question of empirical under-determination.

\subsection{Empirical equivalence}\label{emeq}

Quine's (1975:~p.~319) criterion of empirical equivalence is syntactic: two theories are {\it empirically equivalent} if they imply the same observational sentences, also called observational conditionals, for all possible observations---present, past, future or `pegged to inaccessible place-times' (p.~234) 

Another influential and, as we will see, complementary account of the meaning of `empirical' is by van Fraassen (1980:~p.~64), who puts it thus: 

\begin{quote}\small To present a theory is... to present certain parts of those models (the {\it empirical substructures}) as candidates for the direct representation of observable phenomena. The structures which can be described in experimental and measurement reports we call {\it appearances}: the theory is empirically adequate if it has some model such that all appearances are isomorphic to empirical substructures of that model. 
\end{quote}
Van Fraassen famously restricts the scope of `observable phenomena' to observation by the unaided human senses. Accordingly, his mention of `experimental and measurement reports' is restricted to certain kinds of experiments and measurements.\footnote{For example, van Fraassen's conception of observability rules out collider experiments and astronomical observations, where the reports are based on computer-generated data that encode observations by artificial devices. A conception of observability that is more straightforwardly applicable to modern physics is in Lenzen (1955).} Thus I will set van Fraassen's notion of observability aside but keep his notion of empirical adequacy as a useful {\it semantic} alternative to Quine's syntactic construal of the empirical---and in this sense the two views are complementary to each other. 

In Section \ref{eed}, I will give a judicious reading of these two criteria of empirical equivalence, which will give us a verdict that duals, i.e.~dual theories, are empirically equivalent, as a surprising but straightforward application of van Fraassen's and Quine's proposals.

\subsection{Theoretical equivalence}\label{Peq}

The second notion entering the definition of under-determination$_0$---namely, {\it logical} equiv\-a\-lence---was replaced by Quine, and later by others, by other (weaker) notions, often under the heading of `theoretical equivalence'. In this Section, I will introduce some of these notions, and then present my own account, following De Haro (2019). 

Note that `logical equivalence' is defined in logic books as relative to a vocabulary or signature, and so it is obviously too strict.\footnote{See, for example, Hodges (1997:~pp.~37-38).} For example, one would not wish to count French and English formulations of the theory of electrodynamics as different theories, while they {\it would} count as logically inequivalent by the criterion in logic books, since their vocabularies are different.

Quine also argues that logical equivalence is too strong a criterion. He proposes the following {\it criterion of equivalence} between theory formulations: `I propose that we count two formulations as formulations of the same theory if, besides being empirically equivalent, the two formulations can be rendered identical by switching predicates in one of them' (Quine, 1975:~p.~320). He broadens this criterion further to allow not only `switchings' of terms, but more generally `reconstrual of predicates', i.e.~any mapping of the lexicon of predicates that is used by the theory, into open formulas (i.e.~mapping $n$-place predicates to $n$-variable formulas). For a formalization of this, see Barrett and Halvorson (2016:~pp.~4-6). I will follow these authors in calling this new kind of theoretical equivalence {\it Quine equivalence}. Thus we arrive at:\\
\\
{\bf Quinean under-determination:} two theory formulations are under-determined if they are empirically equivalent but there is no reconstrual of predicates that renders them logically equivalent (i.e.~they are Quine-inequivalent).\\

Barrett and Halvorson (2016:~pp.~6-8) argue that Quine equivalence is too liberal a criterion for theoretical equivalence (which, in turn, means that one ends up with a strict notion of empirical under-determination, which will have few instantiations). Indeed, they give several examples of theories that are equivalent according to Quine's criterion, but that one has good reason to consider inequivalent. They then introduce another criterion, due to Glymour (1970), that better captures what one means by {\it intertranslatability,} i.e.~the existence of a suitable translation between two theory formulations. The criterion is, roughly speaking, the existence of reconstrual maps ``both ways'', i.e.~from $T$ to $T'$ and back. %, when these theories are formulated in first-order logic. 
They show that, in first order logic, intertranslatability is equivalent to the notion of {\it definitional equivalence}: which had already been defined in logic and advocated for philosophy of science by Glymour (1970:~p.~279). Since the criterion of intertranslatability is Quine's criterion taken ``both ways'', it can be seen as an improvement of it. 

%Quine is explicit that he wishes to avoid problems of the meanings of terms, i.e.~his criterion of reconstrual of predicates is formal %: `I want to preserve [the man in the street's insight that $T$ and $T'$ are two formulations of the same theory] while avoiding problems raised by the terms `meaning' and `translation''
%(Quine, 1975:~p.~320). For, as we know, Quine was sceptical about meaning and ontology in general:
%according to Quine (1960:~\S2), externally expressed language always has a {\it referential indeterminacy.}
% according to which the linguist, upon hearing the native utter the word `gavagai' while pointing at a rabbit, might translate it as `rabbit'---while still being at a loss whether to construe the objects as rabbits, or rabbit-stages, i.e.~brief temporal parts of rabbits, or mereological fusions of spatial parts of rabbits. 

%Thus Barrett and Halvorson (2016) seem to follow Quine's lead by making no mention of the meaning of words or interpretation (other than interpretation through formulas in a formal language). They also do not mention a criterion of empirical equivalence as a requirement for theoretical equivalence. Likewise, Weatherall (2015) and Barrett (2018) do not dwell on matters of meaning, interpretation or ontology. Weatherall does mention empirical equivalence---but not full interpretative equivalence---as a necessary condition for theoretical equivalence. (See also Hudetz (2019).)

In De Haro (2019), 
%I dubbed this a {\it quietist} position, held by these authors engaged with theoretical equivalence, and for whom interpretative equivalence is a minimal requirement that their project needs, but on which they do not wish to focus. And this position is tenable and sensible (as long as it is not taken as a {\it replacement} of the interpretative project), i.e.~
I argued that there is an interesting project of finding criteria of equivalence that are mostly formal, %. I dubbed the formal project `weak theoretical equivalence', 
while the {\it full} project, of formal plus interpretative equivalence that I am interested in here, 
%is called `theoretical equivalence'. Indeed my own position is that theoretical equivalence does not require just formal equivalence, but also 
requires the consideration of ontological matters. In particular, theoretical equivalence requires that the interpretations are the same. Thus we can give the following definition:\\
\\
{\bf Theoretical equivalence:} two theory formulations are theoretically equivalent if they are formally equivalent and, in addition, they have the same interpretations.\\
\\
%I have dubbed the project of theoretical equivalence, in the context of the physical sciences (and especially as motivated by dualities), {\it physical equivalence.} 
The following Sections will further articulate the above definition. % by adding three things: (i) a conception of formal equivalence in terms of duality (Section \ref{eed}), (ii) a clarification of the notion of `same interpretation' through the conception of a domain of application, (iii) a discussion of the epistemic status of theoretical or physical equivalence, in particular under what conditions one is justified in making a judgment of theoretical equivalence (both (ii) and (iii) in Section \ref{secdual}). 
Thus {\it pace} Quine, the criterion of individuation of theories that is relevant to scientific theories is not merely formal, but is a criterion of theoretical equivalence. Interpretation matters in science,\footnote{See e.g.~Coffey (2014).} and two theories can only be said to be equivalent if they have the same interpretations, i.e.~they have the same ontology. Thus taking theoretical equivalence as our notion of equivalence, we arrive at the following conception of under-determination, based on Quine's original notion, but now with the correct criterion of individuation of physical theories:\\
\\
{\bf Empirical under-determination:} two theory formulations are under-determined if they are empirically equivalent but theoretically inequivalent.\\
\\
As we will see in the next Section, the most straightforward way to look for {\it examples} of empirical under-determination is if the theories have {\it different interpretations}. This is because, as I argued above, empirical under-determination is primarily a matter of meaning, interpretation, and ontology. 

In what follows, we will always deal with theory formulations, and so will not need to distinguish between `theory' and `theory formulation'. So for brevity, I will from now on often talk of `theories' instead of `theory formulations'.

\section{Duality}\label{secdual}

This Section introduces the notion of duality, along with our Schema (with Butterfield) for understanding it, and a few examples.\footnote{The full Schema is presented in De Haro (2020), De Haro and Butterfield (2017), Butterfield (2020), and De Haro (2019). Further philosophical work on dualities is in  Rickles (2011, 2017), Dieks et al.~(2015), Read (2016), De Haro (2017), Huggett (2017), Read and M\o ller-Nielsen (2020). See also the special issue Castellani and Rickles (2017).} 
The Schema encompasses our overall treatment of dualities: which comprises the notions of bare theory, interpretation, duality, and theoretical equivalence and its conditions (Section \ref{DSchema}); and two different kinds of interpretations, dubbed `internal' and `external' (Section \ref{iei}). Section \ref{exdu} discusses two examples of dualities.

\subsection{The Schema for dualities}\label{DSchema}

In this Section, I illustrate the Schema for dualities with an example from elementary quantum mechanics. 

Consider position-momentum duality in one-dimensional quantum mechanics, represented on wave-functions by the Fourier transformation. For every position wave-function with value $\psi(x)$, there is an associated momentum wave-function whose value is denoted by $\ti\psi(p)$, and the two are related by the Fourier transformation as follows:
\bea\label{Fourier}
\psi(x)&=&{1\over\sqrt{2\pi\hbar}}\int_{-\infty}^\infty\dd p~\ti\psi(p)~e^{{i\over\hbar}\,x\,p}\\
\ti\psi(p)&=&{1\over\sqrt{2\pi\hbar}}\int_{-\infty}^\infty\dd x~ \psi(x)~e^{-{i\over\hbar}\,x\,p}~.\nonumber
\eea
Likewise for operators: any operator can be written down in a position representation, $A$, or in a momentum representation, $\ti A$. Because of the linearity of the Fourier transformation, all the transition amplitudes are invariant under it, so that the following holds (using standard textbook bra-ket notation for the inner product):
\bea\label{matrixe}
\bra \psi|A|\psi'\ket=\bra\ti\psi|\ti A|\ti\psi'\ket~.
\eea
Since the Schr\"odinger equation is linear in the Hamiltonian and in the wave-function, the equation can itself be written down and solved in either representation. 

The Fourier transformation between position and momentum gives us an isomorphism between the position and momentum formulations of quantum mechanics. 
For Eq.~\eq{Fourier} maps, one-to-one, each wave-function in the position formulation to the corresponding wave-function in the momentum formulation, and likewise for operators. Furthermore, the map preserves the dynamics, i.e.~the Schr\"odinger equation in position or in momentum representation, and all the values of all the physical quantities, i.e.~the transition amplitudes Eq.~\eq{matrixe}, are invariant under this map. 

Position-momentum duality illustrates what we mean by duality in general: a {\it duality} is an isomorphism between theories, which I will denote by $d:T\rightarrow T'$. States, quantities, and dynamics are mapped onto each other one-to-one by the duality map, while the values of the quantities, Eq.~\eq{matrixe}, which determine what will be the physical predictions of the theory, are invariant under the map.

Notice the dependent clause `which determine {\it what will be} the physical predictions of the theory'. The reason for the insertion of the italicised phrase is, of course, that my talk of quantum mechanics has so far been mostly formal. Unless one specifies a physical interpretation for the operators and wave-functions in Eq.~\eq{matrixe}, the formalism of quantum mechanics does not make any physical predictions at all, i.e.~only {\it interpreted} theories make physical predictions. Without an interpretation, the formalism of quantum mechanics could equally well be describing some probabilistic system that happens to obey a law whose resemblance with Schr\"odinger's equation is only formal.

This prompts the notions of {\it bare theory} and of {\it interpretation} which, together, form what we call a `theory'. A bare theory is a theory (formulation) before it is given an interpretation---like the formal quantum mechanics above: it was defined in terms of its states (the set of wave-functions), quantities (operators) and dynamics (the Schr\"odinger equation). Thus it is useful to think of a bare theory in general as a triple of states, quantities, and dynamics. I will usually denote bare theories by $T$, and a duality will be an isomorphism between {\it bare theories}.\footnote{One may object that the position and momentum representations of quantum mechanics are not two different bare theories, but two different formulations of the same theory. But recall that I have adopted `theory' instead of `theory formulation' for simplicity, so that it is correct to consider a duality relating two formulations of the same theory.} % There is no requirement that a duality must relate two bare theories that are judged to be distinct.}

Interpretations can be modelled using the idea of {\it interpretation maps}. Such a map is a structure-preserving partial function mapping a bare theory (paradigmatically: the states and the quantities) into the theory's domain of application, i.e.~appropriate objects, properties, and relations in the physical world. I will denote such a (set of) map(s) by: $i:T\rightarrow D$.\footnote{If a bare theory is presented as a triple of states, quantities, and dynamics, then the interpretation is a triple of maps, on each of the factors. However, I will here gloss over these details: for a detailed exposition of interpretations as maps, the conditions they satisfy, and how this formulation uses referential semantics and intensional semantics, see De Haro (2019, 2020) and De Haro and Butterfield (2017).\label{triplei}}

The above notions of bare theory, of interpretation as a map, and of duality as an isomorphism between bare theories, allow us to make more precise the notion of theoretical equivalence, from Section \ref{Peq}. First, suppose that we have two bare theories, $T_1$ and $T_2$, with their respective interpretation maps, $i_1$ and $i_2$, which map the two theories into their respective domains of application, $D_1$ and $D_2$. Further, assume that there is a duality map between the two bare theories, $d:T_1\rightarrow T_2$, i.e.~an isomorphism as just defined. {\it Theoretical equivalence} can then be defined as the condition that the domains of application of the two theories are the same, i.e.~$D:=D_1=D_2$, as in Figure \ref{dual3}, so that the two interpretation maps have the same range. Note that, while the domains of application of theoretically equivalent theories are the same, the bare theories $T_1$ and $T_2$ are different, and so the theories are {\it equivalent} but not {\it identical}.
\begin{figure}
\begin{center}
\bea
\begin{array}{ccc}T_1\!\!&\overset{d}{\longleftrightarrow}&\!T_2\\
~~~~~~~{\sm{$i$}}_{\tn{1}}\searrow&&\swarrow \sm{$i$}_{\tn{2}}~~~~~~~~\\
&D&\end{array}\nonumber
\eea
\caption{\small Theoretical equivalence. The two interpretations describe ``the same sector of reality'', so that the ranges of the interpretations coincide.}
\label{dual3}
\end{center}
\end{figure}

As I mentioned in the definition of empirical under-determination in Section \ref{Peq}, in this paper we are mostly interested in situations in which the two theories are theoretically {\it in}equivalent, i.e.~the ontologies of two dual theories are different. This means that the ranges of the interpretation maps, i.e.~the domains of application of the two theories, are distinct, so that: $D_1\not=D_2$. Thus the diagram for the three maps, $d,i_1,i_2$, does not close: the diagram for theoretical inequivalence is a square, as in Figure \ref{eqmapI}. This notion of theoretical inequivalence makes precise the first condition for under-determination, at the end of Section \ref{Peq}. 

This will allow us, in Section \ref{DandU}, to give precise verdicts of under-determination. In order to do that, we also need to make more precise the second condition for under-determination in Section \ref{Peq}, i.e.~the notion of empirical equivalence. We turn to this next.
\begin{figure}
\begin{center}
\bea
\begin{array}{ccc}T_1&\overset{d}{\longleftrightarrow}&T_2\\
~~\Big\downarrow {\sm{$i_1$}}&&~~\Big\downarrow {\sm{$i_2$}}\\
D_1&\not=&D_2
\end{array}\nonumber
\eea
\caption{\small Theoretical inequivalence. The two interpretations describe ``different sectors of reality'', so that the ranges of the interpretations differ.}
\label{eqmapI}
\end{center}
\end{figure}

To illustrate the notion of empirical equivalence, let me first briefly discuss the standard textbook interpretation of quantum mechanics, in the language of the Schema---the other interpretations are obtained by making the appropriate modifications. Since my aim here is only to illustrate the Schema, rather than to try to shed light on quantum mechanics itself, I will set aside the measurement problem, and simply adopt the standard Born rule for making probabilistic predictions about the outcomes of measurements. On this understanding, the interpretation map(s) $i$ map the bare theory to its domain of application\footnote{See for example van Fraassen (1970:~pp.~329, 334-335).} as follows:
\bea
i(x)&=&\mbox{\small`the position, with value $x$, of the particle upon measurement'}\nn
i\left(|\psi(x)|^2\right)&=&\mbox{\small`the probability density of finding the particle at position $x$, upon measurement'}\nn
i(\psi)&=&\mbox{\small`the physical state of the system (the particle)'},\nonumber
\eea
etc., where $x$ is an eigenvalue of the position operator.\footnote{Note that there are different kinds of elements that are here being mapped to in the domain. The first example is a standard extensional map, which gives a truth value to an observational conditional in a straightforward way. The second maps to the probability of an outcome over measurements of many identically prepared systems, rather than measurement of a single system. The third map describes the physical situation of a particle with given properties: thus the map's range is a concrete rather than an abstract object. \label{intensionals}}

Although all of the above notions (outcomes of measurements, Born probabilities, physical states of a system, etc.) are in the domain $D$, and they all enter in the assessment of theoretical equivalence, not all of them are relevant for empirical equivalence, since not all of them qualify as `observable', in the sense of Section \ref{emeq}. For while the position of a particle can be known by a measurement of the particle's position on a detection screen or in a bubble chamber, the state of a particle cannot be so known: thus it is a theoretical concept. On the other hand, a probability (on the standard---and admittedly simplistic!---frequentist interpretation of probabilities in quantum mechanics, where probabilities are relative frequencies of events for an ensemble of identically prepared systems) is an observational concept, since it can be linked to relations between measurements (i.e.~frequencies), even if it does not itself correspond to an outcome of a measurement or physical interaction.

Thus in the case of quantum mechanics, the {\it empirical substructures} in van Fraassen's definition in Section \ref{emeq} are the subsets of structures of the theory that correspond to outcomes of measurements, in a broad sense, and relations between measurements. These structures are, quite generally, the set of (absolute values of) transition amplitudes of self-adjoint operators, Eq.~\eq{matrixe}, and expressions constructed from them, including powers of transition amplitudes. These are the empirical substructres (on van Fraassen's semantic conception), which make true or false the observational conditionals of the theory (Quine's phrase). The remaining terms of the theory may have a physical significance, but they are theoretical. Thus this discussion distinguishes, for elementary textbook quantum mechanics, between what is theoretical and what is empirical.

\subsection{Examples of dualities}\label{exdu}

I will now give two other important examples of dualities: T-duality in string theory, and black hole-quark-gluon plasma duality. 

To this end, I will first say a few words about string theory. String theory is a generalisation of the theory of relativistic point particles to relativistic, one-dimensional extended objects moving in time, i.e.~strings. On analogy with the world-line swept out by a point particle, the two-dimensional (one space plus one time dimension!) surface  swept out by the string in spacetime is called the `world-sheet' of the string. Strings can be open or closed, so that their world-sheet is topologically an infinitely long strip (for open strings) or an infinitely long tube (for closed strings). Type I theories contain both open and closed strings, while type II theories contain only closed strings. In what follows, I will focus on type II theories. In order to render string theories quantum mechanically consistent, fermionic string excitations are introduced that make the theories supersymmetric, i.e.~so that the number of bosonic degrees of freedom is matched by the number of fermionic degrees of freedom. Depending on the chirality of these fermionic excitations, i.e.~roughly whether they are all handed in the same direction on the world-sheet or not, we have a type IIA (left-right symmetric) or a type IIB theory (left-right antisymmetric). The quantum mechanical consistency of string theory also requires that the number of spacetime dimensions be 10, so that the spacetime is the Lorentzian $\mathbb{R}^{10}$.

T-duality first appears when one of these ten dimensions is compact, for example a circle of radius $R$.\footnote{For an early review, see Schwarz (1992). See also Huggett (2017), Read (2016), Butterfield (2020). I follow the physics convention of setting $\hbar=1$.} Thus the spacetime is $\mathbb{R}^9\times S^1$, where $S^1$ is the circle. The periodicity of the circle then entails that the centre-of-mass momentum of a string along this direction is quantised in units of the radius $R$; so $p=n/R$, where $n$ is an integer. Furthermore, closed strings can wind around the circle (this is only possible if the spacetime is $\mathbb{R}^9\times S^1$ not $\mathbb{R}^{10}$): so they have, in addition to momentum, an additional winding quantum number, $m$, which counts the number of times that a string wraps around the circle. Now the contribution of the centre-of-mass momentum, and of the winding of the string around the circle, to the square of the mass $M^2$ of the string, is quadratic in the quantum numbers, as follows:
\bea\label{mass}
M^2_{nm}={n^2\over R^2}+{m^2R^2\over\ell_{\sm s}^4}~,
\eea
where the subscript $nm$ indicates that this is the contribution of the momentum and the winding around the circle---there are other contributions to the mass that are independent of the momentum, winding, and radius, which I suppress---and $\ell_{\sm s}$ is the fundamental string length, i.e.~the length scale with respect to which one measures distances on the world-sheet. We see that the contribution to the mass, Eq.~\eq{mass}, is invariant under the simultaneous exchange:
\bea\label{Tmn}
(n,m)&\leftrightarrow&(m,n)\\
R&\leftrightarrow&\ell^2_{\sm s}/R~.\nonumber
\eea
In fact, one can show that the entire spectrum, and not only the centre-of-mass contribution Eq.~\eq{mass}, is invariant under this map (cf.~Zwiebach (2009:~pp.~392-397), Polchinski (2005:~pp.~235-247)). This is the basic statement of T-duality: namely, that the theory is invariant under Eq.~\eq{Tmn}, i.e.~the exchange of the momentum and winding quantum numbers, in addition to the inversion of the radius.\footnote{Taking into account how T-duality acts on the fermionic string excitations, one can show that it exchanges the type IIA and type IIB string theories compactified on a circle.}

T-duality is a duality in the sense of the previous Section, in that it maps states of type IIA string theory to states of type IIB, and likewise for quantities, while leaving all the values of the physical quantities, such as the mass spectrum Eq.~\eq{mass}, invariant.\\

My second example is gauge-gravity duality, itself a large subject (cf.~De Haro, Mayerson, Butterfield, 2016), but here I focus on its use in the Relativistic Heavy Ion Collider (RHIC) experiments carried out in Brookhaven, NY. Here, the duality successfully relates the four-dimensional quantum field theory (QCD, quantum chromodynamics) that describes the quark-gluon plasma, produced in high-energy collisions between lead atoms, to the properties of a five-dimensional black hole. The latter was employed to perform a calculation that, via an approximate duality, provided a result in QCD: namely, the shear-viscosity-to-entropy-density ratio of the plasma, which could not be obtained in the theory of QCD  describing the plasma. Thus a five-dimensional black hole was used to describe, at least approximately, an entirely different (four-dimensional!) empirical situation. \\

The level of rigour to which duality has been established differs among various examples. T-duality can be proven perturbatively, i.e.~order by order in perturbation theory; but not (yet) exactly. Gauge-gravity duality is hard to prove beyond the semi-classical approximation: and, although much evidence has been amassed, it is still a conjecture. A rigorously proven duality is Kramers-Wannier duality (see e.g.~Baxter (1982)): this, and other cases treated elsewhere,\footnote{Another example of a mathematically proven duality is bosonization, or the duality between bosons and fermions in two dimensions. See De Haro and Butterfield (2017:~Sections 4 and 5). Cf.~also Butterfield (2020).} show that nothing that I will have to say depends on the conjectural status of the string theory dualities. Indeed, my analysis here will be independent of those details.

\subsection{Internal and external interpretations}\label{iei}

In the previous two Sections, I defined dualities formally (Section \ref{DSchema}) and then I gave two examples (Section \ref{exdu}). The examples already came with an interpretation, i.e.~I discussed not just the bare theories but the interpreted theories, since that is the best way to convey the relevant physics. In this Section, I will be more explicit about the kinds of interpretations that can lead to theoretically equivalent theories.

Not all interpretations lead to theoretically equivalent theories: and this fact---that theoretical equivalence is not automatic---creates space for empirical under-determination. For, as I will argue in the next Section, the interpretations that {\it fail} to lead to theoretical equivalence introduce the possibility of empirical under-determination. 

Which interpretations lead to theoretically equivalent theories, as in Figure \ref{dual3}, or to theoretically inequivalent ones, as in Figure \ref{eqmapI}? Note that theoretical equivalence requires that the domains of the two theories (the ranges of the interpretation maps) are the same, i.e.~the two theories have the same ontology.\footnote{As I stress in this Section, duality does not automatically give theoretical equivalence, because dual theories can have different ontologies. %At the end of this Section, I summarise previous work, in De Haro (2016, 2019), about the conditions under which one is justified in taking duals to be theoretically equivalent, so that dual theory formulations have the same ontology and count as being the {\it same theory}.
} This imposes a strong condition on the interpretations of two dual theories: whereas theoretical inequivalence comes at a low cost. Thus one expects that, generally speaking, an interpretation of two dual theories---for example, the interpretations with which dual theories have been historically endowed---renders two theories inequivalent. I will call such interpretations, that deliver theoretically inequivalent theories, {\it external}. 

An {\it external} interpretation is best defined in contrast with an {\it internal} interpretation, which maps all of and only what is common to the two theory formulations, i.e.~an internal interpretation interprets only the invariant content under the duality. An external interpretation, by contrast, also interprets the content that is not invariant under the duality---thereby typically rendering two duals theoretically, and potentially also empirically, inequivalent.

I will dub this additional structure, which is not part of what I have called the `invariant content', the {\it specific structure.} An isomorphism of theories maps the states, quantities, and dynamics, but not the specific structure, which is specified additionally for each specific theory formulation. Indeed, theory formulations often contain structure (e.g.~gauge-dependent structure) beyond the bare theory's empirical and theoretical structure (which is gauge-independent). %For example, in the case of quantum mechanics that I discussed in Section \ref{DSchema}, the position and momentum representations differ in the choice of variables $x$ and $p$ on which their respective wave-functions depend, i.e.~whether they are written in the position or momentum representations. On the other hand, the states in the Hilbert space and e.g.~the values of the quantities are of course independent of the representation. 

In the example of T-duality, the interpretation of string quantum numbers as `momentum' or as `winding' is {\it external} and makes two T-dual theories inequivalent, since the interpretation maps to distinct elements of the domain under the exchange in Eq.~\eq{Tmn}.

If an interpretation does not map the specific structure of a theory but only common structure, I will call it an {\it internal interpretation}. This means that it only maps the structure that is common to the duals. Such an interpretation gives rise to two maps, $i_1$ and $i_2$, one for each theory: their domains differ, but their range can be taken to be the same, since for each interpretation map of one theory there is always a corresponding interpretation map of the other theory with the same domain of application.\footnote{This conception gives a formal generalisation of an earlier characterisation, in Dieks et al.~(2015:~pp.~209-210) and De Haro (2020:~Section 1.3), of an internal interpretation as one where the meaning of the symbols is derived from their place and relation with other symbols in the theoretical structure, i.e.~not determined from the outside. Those papers were concerned with theory construction, the idea being that when, in order to interpret $T$'s symbols, we e.g.~couple a theory $T$ to a theory of measurement, we do this either through $T$'s specific structure, or by changing $T$ in some other way. Either way, we have an external interpretation, because the specific structure makes an empirical difference. Thus an internal interpretation, which does not introduce an external context or couplings, only concerns facts internal to the triples and our use of them.} Thus we get the situation of theoretical equivalence, in Figure \ref{dual3}.\footnote{The phrase `the internal interpretation does not map the specific structure' can be weakened: one can allow that the interpretation map maps the specific structure to the domain of application, as long as the duality is respected (i.e.~the domains of two duals are still the same), and the empirical substructures of the domain do not change.\label{weakenII}}\\
%: their domains differ, but their range is the same, since by definition they have the same domain of application.
\\
{\bf Unextendability justifies the use of an internal interpretation.} %So far we have discussed internal interpretations as those that do not map the specific structure of theories. %, so that two dual theories have the same domain of application under such an interpretation: the two theories are theoretically equivalent. %External interpretations are all the others, i.e.~those that also map the specific structure, and so they can distinguish dual models, and map them to different domains of application. 
%However, t
There is an important epistemic question, that will play a role in Section \ref{physin}'s analysis, about whether we are {\it justified} in adopting an internal interpretation, according to which the duals are theoretically equivalent. 

The question is, roughly speaking, whether the interpretation $i$ of a bare theory $T$, and $T$ itself, are ``detailed enough'' and ``general enough'' that they cannot be expected to change, for a given domain of application $D$. I call this condition {\it unextendability.} %and it is a condition on the interpreted theory, i.e.~the bare theory plus its interpretation. 
De Haro (2020:~Section 1.3.3) gives a technical definition of the two conditions for unextendability, the `detail' and `generality' conditions. Roughly speaking, the condition that the interpreted theory is detailed enough means that it describes the entire domain of application $D$, i.e.~it does not leave out any details. And the generality condition means that both the bare theory $T$ and the domain of application $D$ are general enough that the theory ``cannot be extended'' to a theory covering a larger set of phenomena. Thus $T$ is as general as it can/should be, and $D$ is an entire ``possible world'', and cannot be extended beyond it. 

To give an example involving symmetries: imagine an effective quantum field theory with a classical symmetry, valid up to some cutoff, and imagine an interpretation that maps symmetry-related states to the very same elements in the domain of application. Now imagine that extending the theory beyond the cutoff reveals that the symmetry is broken (for example, by a higher-loop effect), so that it is anomalous. The possibility of this extension, with the corresponding breaking of the symmetry, will lead us to question that our interpretation is correct: maybe we should not identify symmetry-related states by assigning them the same interpretation, especially if it turns out that the states are different after the extension (for example, if higher-loop effects correct the members of a pair of symmetry-related states in different ways). And so, it will probably prompt us to develop an interpretation that is consistent with the theory's extension to high energies, so that symmetry-related states are now mapped to distinct elements in the domain. In conclusion, we are not epistemically justified in interpreting symmetry-related states as equivalent, because we should take into account the possibility that an extension of the theory to higher energies might compel us to change our interpretation.

External interpretations of two dual theories do not in general classify them as having the same interpretation. And so, although these interpretations could be subject to change, this does not affect the verdict that the theories are theoretically inequivalent. 

%Notice that, for a theory to be unextendable, it is not enough for it to describe an entire possible world in full detail. One needs, in addition, some argument to the effect that the theory is in some sense unique or, better, isolated in the space of related theories, and that in that sense the theory cannot be extended. For if it could be extended, the interpretation could thereby map to a different possible world, thus invalidating the former interpretation and hence the verdict of theoretical equivalence. As I will argue in a moment, symmetries can secure unextendability, since in effect they give arguments that a theory is unique within a set of theories based on a given set of fields, and requiring a given set of symmetries.

There is a weaker sense of unextendability, which allows that a theory might be extended in some cases, but only in such a way that the interpretation in the original domain of application does not change, and the theories continue to be theoretically equivalent after the extension. In what follows, I will often use `unextendability' in this weaker sense.

The example of position-momentum duality in elementary quantum mechanics can be interpreted internally once we have developed quantum mechanics on a Hilbert space. The two theories are then representations of a single Hilbert space, and we can describe the very same phenomena, regardless of whether we use the momentum or the position representation. And quantum mechanics is an unextendable theory in the weaker sense (namely, with respect to this aspect with which we are now concerned, of position vs.~momentum interpretations) because, even if we do not have a ``final'' Hamiltonian (we can often add new terms to it), the position and momentum representations keep their power of describing all possible phenomena equally well. Namely, because of quantum mechanics' linear and adjoint structure, unitary transformations remain symmetries of the theory regardless of which terms we may add to the Hamiltonian: so that our interpretation in terms of position or momentum will not change.

Dualities in string and M-theory are also expected to be dualities between unextendable theories, at least in the weak sense. For example, the type II theories discussed in Section \ref{exdu} have the maximal amount of supersymmetry possible in ten dimensions; and the number of spacetime dimensions is determined by requiring that the quantum theories be consistent. Also, their interactions are fixed by the field content and the symmetries. If we imagine, for a moment, that these theories are exactly well-defined (since the expectation is that M-theory gives an exact definition of these theories, and that T-duality is a manifestation of some symmetry of M-theory), then they are in some sense ``unique'', constrained by symmetries, and thus unextendable: they are picked out by the field content, their set of symmetries, and the number of spacetime dimensions.\footnote{See also Dawid (2006:~pp.~310-311), who discusses a related phenomenon of `structural uniqueness'. } Thus, if these conjectures can be fleshed out, taking type IIA and type IIB on a circle to be theoretically equivalent will be justified, because the theories describe an entire possible world, and there is no other theory ``in their vicinity''. 

\section{Duality and Under-determination}\label{DandU}

In this Section, I first make some remarks about aspects of scientific realism relevant for dualities (Section \ref{SciRe}). Then I will illustrate the notions of theoretical {\it inequivalence} of dual theory formulations (Section \ref{physin}) and empirical equivalence (Section \ref{eed}). %This includes the consideration of a judicious reading of the notion of empirical equivalence from Section \ref{emeq}. %, according to an analysis of the semantic and syntactic conceptions of theories. 
This 
%judicious reading 
will lead to a surprising but straightforward conclusion, which will give us our final verdict about whether dualities admit empirical under-determination. 
Section \ref{abstr} discusses how to obtain internal interpretations. Section \ref{trouble} will then ask whether scientific realism is in trouble. 

\subsection{Scientific realism: caution and-or virtue?}\label{SciRe}

My discussion of scientific realism aims to be as general as possible, i.e.~as independent as possible of a specific scientific realist position.\footnote{My own position is in De Haro (2020a:~pp.~27-59).} %However, let me specify a bit more. 
The notion of under-determination that we are considering has both semantic and epistemic aspects (see Section \ref{Qbeudt}), and also the interesting scientific realism to consider has both semantic and epistemic aspects. Roughly, the relevant scientific realism is the belief in the approximate truth of scientific theories that are well-confirmed in a given domain of application.

%`Truth' is here understood semantically, say in terms of the truth values of sentences. However, this 
The semantic aspect does not involve a na\"ive ``direct reading'' of a scientific theory (as some formulations, like van Fraassen's (1980), could lead us to think). Rather, it involves a literal, but nevertheless {\it cautious,} reading, informed by current scientific practice and by history. This is essential to secure that the belief in the theory's statements is epistemically justified. Indeed, one does not simply take the nearest scientific textbook and quantify existentially over the entities that are defined on the page. There are many cases where one is not justified in believing in the entities that are introduced by even our best scientific theories, and so one proceeds tentatively. In such cases, a {\it cautious} scientific realist ought to suspend judgment about the existence of the posited entities, until further interpretative work has been carried out that justifies the corresponding belief. For example, consider the cases of local gauge symmetries, and of a complex phase in the overall wave-function of a system: in both cases, belief in the corresponding entities ought to be postponed until the further analysis is done. This contrasts with, say, the particle content of our best theories, whose confirmed existence, under normal circumstances, one accepts (even if the particles are microscopic).

Proponents of selective scientific realist views, for example Psillos (1999) and Kitcher (1993), are in this sense cautious. Their accounts assign reference only to those entities that are in some sense indispensable or important to explain the empirical success of the theories, while stating that other terms (like `phlogiston') were less central to the theories' success, e.g.~because they are not causally involved in the production of the relevant phenomena, and so should be regarded as non-referring, or as referring on some occasions but not others.\footnote{These accounts have been criticised on various points, most notably because the defence of selective confirmation appears to involve a selective reading of the historical record (Stanford, 2006:~p.~174).}

While I do not myself endorse the details of these proposals, I am sympathetic to them: and they {\it do} illustrate the general idea of `caution' about realist commitments endorsed above. Namely, that determining the realist's commitments is not a matter of ``reading off'', but sometimes involves judgment. 

Let me now say more about how this applies to dualities (more details in Section \ref{trouble}), without aiming to develop a new scientific realist view here: since, as I said, my arguments should apply to different versions of scientific realism. Rather, I wish to express a general attitude---which I denote with the word `caution'---towards the role of inter-theoretic relations in the constitution of scientific theories.

In so far as a duality can have semantic implications---namely, in so far as dualities contribute to the criteria of theory individuation---the cautious scientific realist should take notice of those implications, and suspend judgment about whether two dual theory-formulations are distinct theories, until those criteria have been clarified. Indeed, the account of dualities that I favour proposes that, under sufficient conditions of:

(1) internal interpretation, 

(2) unextendability, 

(3) having a philosophical conception of `interpretation',\footnote{This is my own version of what Read and M\o ller-Nielsen (2020:~p.~266) call `an explication of the shared ontology of two duals'. See De Haro (2019:~Section 2.3.1) and footnote \ref{philcon}.} 

\noindent one is justified (but not obliged) to view duals as notational variants of a single theory. 

This view of theory individuation takes a middle way---``the best of both worlds''---between two positions that, in the recent literature, have been presented as antagonistic. Read and M\o ller-Nielsen (2020:~p.~276) defend what they call a `cautious motivationalism': duality-related models may only be regarded as being theoretically equivalent once an interpretation affording a coherent explication of their common ontology is provided, and there is no guarantee that such an interpretation exists. 

Huggett and W\"uthrich (2020:~p.~19) dub this position an `agnosticism about equivalence' and object that, though it is {\it cautious,} it is less epistemically {\it virtuous} than their own position: namely, to assert that `string theory is promising as a complete unified theory in its domain, and so it is reasonably thought to be unextendable. And from that we do think physical equivalence is the reasonable conclusion'. 

The position that I defend takes a middle way between these apparently contrasting positions: one is justified, but not obliged, (i) to take duals to be theoretically equivalent (under the three conditions, (1)-(3), mentioned above), and (ii) to be a realist about their common core.\footnote{This notion appears to be close to Ney's (2012:~p.~61) `core metaphysics', which retains those common `entities, structures, and principles in which we come to believe as a result of what is found to be indispensable to the formulation of our physical theories'. However, one should keep in mind that I am here concerned with semantics and epistemology, and not chiefly with metaphysics. I thank an anonymous reviewer for pointing out this similarity.}

But even if the three conditions above, (1)-(3), are not fully met, or are not explicitly formulated, it may still be {\it legitimate,} lacking full epistemic justification, to take the duals as equivalent and to be a realist about their common entities: as a working assumption, a methodological heuristic, or a starting point for the formulation of a new theory. %As things stand, only the former option is relevant for the discussion of scientific realism that I am conducting here; the latter option does not meet the epistemic criterion. 

\subsection{Theoretically inequivalent duals}\label{physin}

In Section \ref{DSchema}, I defined theoretically equivalent theories in terms of the triangular diagram in Figure \ref{dual3}, and theoretically inequivalent theories in terms of a diagram that does not close, i.e.~a square diagram as in Figure \ref{eqmapI}.

In Section \ref{iei}, we made a distinction between internal and external interpretations. Thus the Schema gives us two cases in which dual theories can {\it prima facie} be theoretically inequivalent. The first is through the adoption of external interpretations. These interpretations lead to the square diagram in Figure \ref{eqmapI}, with different domains of application. I will discuss this case in Section \ref{trouble}.

The second case is that of two theoretically equivalent but extendable dual theories. %\footnote{There is a third possible case, of internal interpretations for unextendable theories with a diagram like that of Figure \ref{eqmapI} that does not close. However, this is like the case of external interpretations, and need not be treated separately; thus in what follows I will treat these interpretations as being effectively external.}  
In this case, although the judgment, based on a given pair of internal interpretations, is that the theories are theoretically equivalent, this judgment is epistemically unjustified because the theory is extendable (as I discussed in Section \ref{iei}). And since the judgment of theoretical equivalence is unjustified, this could {\it prima facie} give a new way to get theoretically inequivalent theories.

However, this second case is not a genuine new possibility. For the lack of justification for adopting the internal interpretations prompts us to either: (i) Justify the use of these, or some other, internal interpretations, thus getting a justified verdict of theoretical equivalence; or (ii) conclude that the internal interpretations under consideration are indeed not adequate interpretations, and adopt external interpretations instead, which then judge the two theories to be {\it inequivalent.}

In case (i), where one finds internal interpretations whose use {\it is} justified, we end up with a case of theoretical equivalence and not inequivalence, i.e.~a case of the triangle diagram Figure \ref{dual3}; and so this case is irrelevant to the under-determination thesis. In case (ii), we are back to the situation of {\it external} interpretations. Thus external interpretations exhaust the interesting options for empirical under-determination.\footnote{My reasoning here bears a formal similarity with Norton's (2006) analysis of Goodman's new riddle of induction. \label{grueify}}

%Could cut out this paragraph:
Recall the examples of external interpretations from Section \ref{iei}. External interpretations of T dual pairs of strings winding on a circle interpret the momentum and winding quantum numbers, $n$ and $m$ respectively, as corresponding to different elements in the domain (namely, to momentum and to winding, respectively), and in addition they would interpret the circle as having a definite radius of either $R$ or $R':=\ell_{\sm s}^2/R$: so that the duality transformation Eq.~\eq{Tmn} leads to physically inequivalent theories, despite the existence of a formal duality that pairs up the states and quantities. Such an external interpretation can for example include ways to measure the quantities of interest---momentum, winding, and the radius of the circle---so that they indeed come to have definite values, according to the external interpretation. An indirect way to measure the radius is by measuring the time that a massless string takes to travel around the circle. 

\subsection{Empirically equivalent duals are under-determined}\label{eed}

In this Section, I will use Section's \ref{emeq} account of empirical equivalence, which will enable our final verdict about empirical under-determination in cases of duality. The Section summarises De Haro (2020b) (see also Weatherall, 2020).

%that the Schema suggests, based on the notions of theory and interpretation from Section \ref{DSchema}. The notion of empirical equivalence is a judicious reading of that from Section \ref{emeq}, where the reading is motivated not only by dualities, but---as I will argue---also by the semantic and syntactic conceptions of theories. This .

%The `judicious reading' of empirical equivalence that I will consider will be liberal (and perhaps surprising) but straightforward. This is partly motivated by the need to give the critic of scientific realism some leeway to produce alternative theories: thus I temporarily play the devil's advocate---but not only. For, as I will discuss at the end of the Section, these alternatives are no playful philosophical fictions.\footnote{I here use `fiction' in the sense of footnotes \ref{stanford} and \ref{bason}.} First, they capture an important scientific practice of using dualities to construct new theories that describe empirically equivalent situations, as in the RHIC experiment. Second, the judicious reading is independently motivated by a historical-critical analysis of van Fraassen's semantic criterion of empirical equivalence. 

According to the syntactic conception of theories, two theories are empirically equivalent if they imply the same observational sentences.\footnote{See Quine (1970, 1975) and Glymour (1970, 1977).} As I argued in Section \ref{iei}, externally interpreted theories are in general {\it not} empirically equivalent, in this sense. The domains are distinct, as Figure \ref{eqmapI} illustrates. % and, in most cases we have seen, these domains even belong to different possible worlds. %For example, T duals which differ in their momentum and winding quantum numbers, and radii $R$ vs.~$R'={\ell_{\sm s}^2/R}$, are clearly not empirically equivalent. For, as I already noticed in the previous Section, a measurement of the time that a photon takes to go around the compact direction gives different results in the two cases. Indeed, we have two different possible worlds, with radii $R$ vs.~$R'$! \\

On the semantic conception, two theories are empirically equivalent if the empirical substructures of their models are isomorphic to each other (cf.~Section \ref{emeq}). 

%As I mentioned in Section \ref{DSchema}, the empirical substructures of the models of a theory can be taken to be the values of various physical quantities evaluated between states, as illustrated in Eq.~\eq{matrixe}, properly interpreted for each of the models. Now van Fraassen's conception of empirical equivalence, from Section \ref{emeq}, says that these empirical substructures are {\it isomorphic} to each other.\footnote{When I mention van Fraassen's conception of empirical equivalence, here and in what follows, I refer to the conception that he puts forward in his (1980, 1989), and summarised in the quote given in Section \ref{emeq}. I submit that this quote should be given a literal reading, and thus that one may actively use isomorphisms, as I do in the main text, to translate between empirically equivalent theories.} Notice that it is `isomorphism' rather than `identity' that counts here. Thus let us look for a suitable {\it isomorphism} between the empirical substructures of the models that we consider. Since we are dealing with dualities, the suggestion is that 

For dualities, the duality map $d$ gives us a natural---even if surprising---new candidate for an isomorphism between the empirical substructures of the models: I will dub it the `induced duality map', $\ti d:D_1\rightarrow D_2$.\footnote{Since we are discussing empirical equivalence, the domains of application can here be restricted to the observable phenomena. However, I will not indicate this explicitly in my notation.} It is an isomorphism between the domains of application, subject to the condition that the resulting (four-map) diagram, in Figure \ref{dtilde}, commutes. This commutation condition is the natural condition for the induced duality map to mesh with the interpretation (the condition for its commutation is that $i_2\,\circ\,d=\ti d\,\circ\,i_1$). %\footnote{For more details, see De Haro (2019:~Section 2.2.3).} 
If such a map exists, then the two theories are clearly {\it empirically equivalent} on van Fraassen's conception, even though they are theoretically inequivalent, because the induced duality map is not the identity, and the domains differ: $D_1\not=D_2$. %\footnote{Since the empirical substructures are substructures of the physical structures, this condition can be weakened. Instead of requiring a map $\ti d:D_1\rightarrow D_2$ between the domains, we can require that only a suitable map between the empirical subsets of the domains closes. The closure condition is then $(i_2\circ d)|_{\tn{emp}}=(\ti d\circ i_1)|_{\tn{emp}}$. \label{emps}}
\begin{figure}
\begin{center}
\bea
\begin{array}{ccc}T_1&\overset{d}{\longleftrightarrow}&T_2\\
~~\Big\downarrow {\sm{$i_1$}}&&~~\Big\downarrow {\sm{$i_2$}}\\
D_1&\overset{\ti d}{\longleftrightarrow}&D_2
\end{array}\nonumber
\eea
\caption{\small Empirical equivalence. There is an induced duality map, $\ti d$, between the domains.}
\label{dtilde}
\end{center}
\end{figure}

Thus, if such an induced duality map exists, on this literal account the dualities that we have discussed {\it do} in fact (and surprisingly!) relate empirically equivalent theories.\footnote{De Haro (2020b) argues that duality is indeed the correct type of isomorphism to be considered, on the semantic criterion of empirical equivalence.} Take for example T-duality: here, by construction, we can map the parts of the theory that depend on the quantum numbers $(n,m)$ and radius $R$ to the quantum numbers $(m,n)$ and radius $R'=\ell^2_{\sm s}/R$, and likewise in the domain, by swapping measurements of positions of momenta, and inverting the physical radii.\\

%Let me add a word about why duality (and induced duality) are the right kinds of isomorphism here, i.e.~why I am allowed to generalise van Fraassen's relatively simple isomorphisms (e.g.~between Newtonian theories that differ only in an overall speed) to more general dualities. To see this, one has to specify what is the relevant `isomorphism', which is a notion that is well-defined only relative to a given structure. In the case of scientific theories, the natural isomorphisms to consider are those that relate their structures. And since van Fraassen (1980, 1989) presents scientific theories in terms of states and quantities, with their corresponding interpretation maps (see also, van Fraassen 1970:~pp.~329, 334-335), the natural isomorphisms should relate these structures. But this is precisely also what the duality and induced duality map do: dualities are maps between the states and the quantities that preserve the values of the quantities (see Section \ref{DSchema}). Thus the claim that dual theories can thus be taken to be empirically equivalent is a natural (although perhaps unexpected!) application of van Fraassen's semantic criterion.\\

%Does the syntactic notion of empirical equivalence, in terms of which discussions of theoretical equivalence have traditionally been framed, allow for the same latitude? I will 
The semantic notion is {\it prima facie more liberal} than Quine's syntactic notion (in the sense that it is less fine-grained, because it gives a verdict of empirical equivalence more easily), and more in consonance with the scientific practice of dualities---thus it prompts a judicious reading of the {\it syntactic} notion of empirical equivalence.
%The syntactic criterion of empirical equivalence can be reinterpreted in a more liberal way, similarly to the semantic criterion (see De Haro, 2020). I
To this end, it is not necessary to change Quine's criterion of empirical equivalence from Section \ref{emeq}; all we need to do is to change one of the theories, generating a new theory by giving a non-standard interpretation to the bare theory. %\footnote{Agreed: it is of course {\it not mandatory} to take dual theories to be empirically equivalent. This is clearest in the syntactic conception, where a theory and its dual can always be interpreted according to their ordinary interpretations, rather than non-standard ones. And the duals then simply {\it disagree} about empirical matters. But also on the semantic view this is possible, by adopting non-isomorphic interpretations.}

Since we have a non-standard and innovative interpretation, %(we have effectively replaced $j|_{\sm{obs}}$ by $t\circ j|_{\sm{obs}}$), 
we have abandoned what may have been the theory's intended meaning, i.e.~we have {\it changed the theory,} stipulating a reinterpretation of its terms, thus producing the desired observational sentences.\footnote{For unextendable theories, the theory's natural interpretation is surely an internal, not an external, interpretation.} %Thus, if we originally thought that the subject-matter of our theory was $S$ (e.g.~a five-dimensional black hole), after the translation we have---to our surprise---a theory whose subject-matter is $t(S)$ (namely, colliding heavy nuclei). 
Thus we have been faithless to the meanings of words. But this is allowed, since we are dealing with external interpretations anyway: and external interpretations can be changed if one's aims change. Indeed, nobody said that we had to stick to a single interpretation of a bare theory in order for it to make empirical predictions: for although theories may have intended interpretations, assigned to them by history and convenience, nothing---more precisely, none of the Quinean notions of empirical under-determination and empirical equivalence---prevents us from generating new theories by reinterpreting the old ones, thus extending the predictive power of a bare theory: but also creating cases of under-determination! %The use of this flexibility in explaining heavy-ion collisions shows that these cases are not contrived or fictitious, but rather scientifically important! 

In Section \ref{abstr}, I first sketch how internal interpretations are usually obtained. For this will be important for the resolution of the problem of under-determination, in Section \ref{trouble}.

\subsection{Obtaining interpretations by abstraction}\label{abstr}

Internal interpretations of dual theory formulations are often obtained by a process of {\it abstraction} from existing (often, historically given) interpretations of the two formulations. Let me denote the {\it common core} theory obtained by abstraction, by $\hat t$, and let $\hat T_1$ and $\hat T_2$ be two empirically equivalent dual theories from which $\hat t$ is obtained. Here, the hats indicate that these are {\it interpreted,} rather than bare, theories. 

We can view the formulation of the common core theory $\hat t$ as a two-step procedure. We first develop the bare i.e.~uninterpreted theory, and then its interpretation: 

(1)~A common core bare theory, $t$, is obtained by a process of abstraction from the two bare theories, $T_1$ and $T_2$, so that these theory formulations are usually representations (in the mathematical sense) of this common core bare theory. This is discussed in detail in De Haro and Butterfield (2017). 

(2)~The internal interpretation of the common core theory, $\hat t$, is similarly obtained through abstraction: namely, by abstracting from the commonalities shared by the interpretations of $\hat T_1$ and $\hat T_2$. In this way, one obtains an internal interpretation that maps only the common core of the two dual theory formulations.\footnote{`Abstraction' does not necessarily mean `crossing out the elements of the ontology that are not common to the two theory formulations'. Sometimes it also means `erasing some of the {\it characterisations} given to the entities, while retaining the {\it number} of entities'. For example, when we go from quantum mechanics described in terms of position, or of momentum, space (Eq.~\eq{Fourier}), to a formulation in terms of an abstract Hilbert space, we do not simply cross out positions and momenta from the theory, but rather make our formulation independent of that choice of basis. In this sense, the formalism of the common core theory is not always a ``reduced'' or ``quotiented'' version of the formalisms the two theories.}

The incompatibility of the interpretations of two theory formulations is thus resolved through an internal interpretation that captures their common aspects, and assigns no theoretical or physical significance to the rest. 

Let me spell out the consequences of this a bit more. The result of the process of abstraction, i.e.~points (1) and (2), is that the interpreted theories $\hat T_1$ and $\hat T_2$ are representations or instantiations of the common core theory, $\hat t$. Thus in particular, we have:
\bea\label{implications}
\hat T_1\Rightarrow\hat t~~~~\mbox{and}~~~~\hat T_2\Rightarrow\hat t~,
\eea
where I temporarily adopt a syntactic construal of theories (but the same idea can also be expressed for semantically construed theories). That is, because $\hat T_1$ and $\hat T_2$ are representations or instantiations of $\hat t$, and in particular because $\hat t$ has a common core interpretation (more on this below): whenever either of $\hat T_1$ and $\hat T_2$ is true, then $\hat t$ is also true.\footnote{Since the sentences of a scientific theory, observable or not, are given by the theory's interpretation, the above entailments should be read as entailments between statements about the world. As I mentioned in (2), this requires that we adopt internal interpretations: so that $\hat t$'s interpretation is {\it obtained from the interpretations} of $\hat T_1$ and $\hat T_2$ by abstraction, and the entailments are indeed preserved.} (For example, think of how a representation of a group, being defined as a homomorphism from the group to some structure, satisfies the abstract axioms that define the group, and thus makes those axioms true for that structure).

Further, for empirically equivalent but theoretically {\it in}equivalent theories, we have:
\bea\label{implication0}
\hat T_1\Rightarrow \neg\,\hat T_2~.
\eea

Let me first give a simple example, not of a full common core theory, but of a single quantity that is represented by different theory formulations. Consider the quantity that is interpreted as the `energy' of a system: while different theory formulations may describe this quantity using different specific structure, and so their external interpretations differ (even greatly) in their details, the basic quantity represented---the energy of the system---is the same, as described by the internal interpretation. Indeed, in all of the following examples, the energy is indeed represented, on both sides of the duality, by quantities that match, i.e.~map to one another under the duality: position-momentum duality (Section \ref{DSchema}), T-duality and gauge-gravity duality (Section \ref{exdu}; Huggett 2017; De Haro, Mayerson, Butterfield, 2016), bosonization (De Haro and Butterfield, 2017), electric-magnetic duality.\\

Read and M\o ller-Nielsen (2020) question that a common core theory $\hat t$ always exists, since they require an explication of the shared ontology of two duals, before a verdict of theoretical equivalence is justified.\footnote{I endorse their requirement, which is more specific than my own earlier requirement of a `deeper analysis of the notion of reference itself' (2020:~Section 1.3.2) and `an agreed philosophical conception of the interpretation' (2019:~Section 2.3.1).\label{philcon}}

Since points (1) and (2) sketch a procedure for obtaining $\hat t$ by abstraction, the question of ``whether $\hat t$ exists'' is effectively the question of whether the common core theory $\hat t$ thus obtained has a well-defined ontology. This can be divided into two further subquestions: 

(1') Is the thus obtained structure $t$, of which $T_1$ and $T_2$ are representations, itself a bare theory, i.e.~a triple of states, quantities, and dynamics? %(cf.~De Haro, 2016:~pp.~259-261). 

(2') Does $t$ also have a well-defined ontology?

If these two questions are answered affirmatively, then $\hat t$ is a common core theory in the appropriate domain of application. 

While examples of dualities have been given in De Haro (2019) for {\it extendable} theories, where the answer to at least one of these questions is negative (so that a common core theory does not exist): I am not aware of any examples of {\it unextendable} theories for which it is known that a common core does not exist.

Let me briefly sketch whether and how some of the familiar examples of dualities satisfy the requirements (1') and (2') (this paragraph is slightly more technical and can be skipped). Consider bosonization (De Haro and Butterfield, 2017). (1'): The set of {\it quantities} is constructed from an infinite set of currents that are the generators of the enveloping algebra of the affine Lie algebra of $\mbox{SU}(2)$ (or other gauge group). The {\it states} are the irreducible unitary representations of this algebra. Finally, the {\it dynamics} is simply given by a specific Hamiltonian operator. (2'): The theory has an appropriate {\it interpretation:} the states can be interpreted as usual field theory states describing fermions and bosons. The operators are likewise interpreted in terms of energy, momentum, etc. And the dynamics is ordinary Hamiltonian dynamics in two dimensions. Thus bosonization answers both (1') and (2') affirmatively.

Other string theory dualities are less well-established, and so the results here are restricted to special cases. In the case of gauge-gravity dualities, under appropriate conditions, the states include a conformal manifold with a conformal class of metrics on it (De Haro, 2020:~p.~278). The quantities are specific sets of operators (which again can be interpreted in terms of energy and momentum), and the dynamics is again a choice of a Hamiltonian operator. Thus, at least within the given idealisations and approximations, and admitting that gauge-gravity dualities are not yet fully understood; they also seem to answer (1') and (2') affirmatively.\footnote{Note that this answer to (2') about the domain of application does not just mean getting correct predictions. % about {\it observable} entities, in the traditional sense. 
For example, common core theories are also being used to give explanations and answer questions that are traditionally regarded as theoretical, e.g.~about locality and causality (Balasubramanian and Kraus, 1999) and about black hole singularities (Festuccia and Liu, 2006). For a variety of other physical questions addressed using the AdS/CFT common core, see Part III of Ammon and Erdmenger (2015).}

%The above discussion assumes that, for unextendable theories, a common core theory $\hat t$ exists that resolves the under-determination. But is this assumption justified? As I already discussed in Section \ref{abstr}, Read and M\o ller-Nielsen (2020) argue that it is not. 

There are two reasons to set aside the worry that $\hat t$ does not exist, together with its threatened problem  of under-determination: so that in Section \ref{trouble} I can safely assume that $\hat t$ exists. (I restrict the discussion to unextendable theories). 
%But I will now argue that this is not an issue, for two reasons. Since the relevant theories for our discussion are unextendable ones, I restrict my discussion to the examples of string theory and quantum field theory. 

First, as just discussed, there is evidence that such common core theories exist in many (not to say all!) examples of such dualities. Indeed, dualities would be much less interesting for physicists if the common core was some arbitrary structure that does not qualify as a physical theory in the appropriate domain of application. Thus it is safe to conjecture, for unextendable theories, that of all the {\it putative} dualities, those that are genuine dualities and are of scientific importance (and that would potentially give rise to a more serious threat of under-determination, because the theory formulations differ from each other more) {\it do} have common cores.

The second, and main, reason is that there are, to my knowledge, no examples of dualities for unextendable theories where a common core is known to {\it not} exist. For some dualities, it is not known whether a common core exists. For example, Huggett and W\"uthrich (2020) mention T-duality as an example where a common core has not been formulated explicitly: working it out explicitly would, in their view, require a formulation of M-theory. It might also be possible to work out a common core theory perturbatively. But so long as it is not known whether a common core exists, rather than having positive knowledge that it does not exist, this just means that there is work to be done: the lack of an appropriate common core theory $\hat t$ can, at best, give a case of {\it transient under-determination,} which, as I argued in Section \ref{eudt}, reflects our current state of knowledge, and is not really worrisome.

This agrees with Stanford's requirement that, before such putative cases of under-determination are accepted, actual examples should be produced (cf.~footnotes \ref{stanford} and \ref{bason}). And so, this worry can be set aside.

%In the case of T-duality, Huggett and W\"uthrich (2020) argue that an explicit common core is {\it not known}: working it out explicitly would, in their view, require a formulation of M-theory. While it might be possible to work out a common core theory perturbatively, I will return to this possibility in Section \ref{trouble}.

\subsection{Trouble for scientific realism?}\label{trouble}

In this Section, I address the question announced at the beginning of Section \ref{physin}: namely, whether, for dual bare theories, the under-determination of {\it external} interpretations gives a problem of empirical under-determination. Let us return to the distinction between extendable and unextendable theories:\\
\\
(A)~{\it Extendable theories:} I will argue that here we have a case of {\it transient under-determi\-na\-tion} (see Section \ref{intro} and footnotes \ref{sofar} and \ref{lyre}), i.e.~under-determination by the evidence so far. And so, I will argue that there is no problem of empirical under-determination here, but only a limitation of our current state of knowledge.

The reason is that extensions of the theory formulations that break the duality are allowed by the external interpretations, i.e.~such that two duals map to a different domain of application with different empirical substructures (or different observation sentences). Interpreting the specific structure, as external interpretations do, introduces elements into the domains that, in general, render the two theory formulations empirically {\it in}equivalent. 

This is in the ordinary business of theory construction. For, although in our current state of knowledge it may appear that the theory formulations are empirically equivalent, the fact that the theory is extendable means that future theory development could well make an empirical difference. % (or else we have unextendable theories in the weak sense). 
Thus one should interpret such theories tentatively. This is the ordinary business of transient under-determination: and one should here look for the ordinary responses of scientific realist positions.\\
\\
(B)~{\it Unextendable theories:} In this case, the theories are somehow ``unique'', perhaps `isolated in the space of related theories' (cf.~Section \ref{iei}). %We should distinguish two cases: \\

%(a) An internal interpretation might not exist, i.e.~there might be no (set of) possible world(s) described by the common core of the theories (e.g.~by taking the common parts of two external interpretations). This possibility is considered by Read and M\o ller-Nielsen (2018). De Haro (2019) gives examples of this for {\it extendable} theories, but I am not aware of any examples for {\it unextendable} theories. Thus I will follow Stanford's requirement that, before this putative under-determination is accepted, actual examples be produced (cf.~footnotes \ref{stanford} and \ref{bason}). And so, this case can be set aside.

%(b) An internal interpretation exists, and so there is under-determination. Here, 

I will argue that a cautious scientific realism (see Section \ref{SciRe}) does not {\it require} belief in either of the  (incompatible) interpretations of the dual theory formulations. Rather, belief in the internal interpretation, obtained by a process of abstraction from the external interpretations, is justified. And so, there is under-determination, but of a kind that is benign. Thus dualities may be taken to favour a cautious approach to scientific realism.

In more detail: as I discussed in Section \ref{SciRe}, in the presence of inter-theoretic relations (and given the three conditions discussed at the end of Section \ref{SciRe}), the cautious scientific realist will take the inter-theoretic relations into account when she determines her realist commitments. Although the under-determination prevents her from being able to choose between $\hat T_1$ and $\hat T_2$, she has an important reason to favour $\hat t$. % over $\hat T_1$ or $\hat T_2$. 
Namely, she will prefer $\hat t$ over $\hat T_1$ and $\hat T_2$ on logical grounds. For, on the basis of the implications Eqs.~\eq{implications} and \eq{implication0}, and regardless of which of $\hat T_1$ and $\hat T_2$ is true, she can consistently (and at this point, also should) accept the truth of $\hat t$. Thus, since her scientific realism commits her to at least one of these three theories (notably, because they are empirically adequate theories, and they otherwise satisfy the requirements of her scientific realism), she will {\it in any case} be commmitted to $\hat t$. Namely, a commitment to either $\hat T_1$ or $\hat T_2$ commits her, in virtue of Eq.~\eq{implications}, {\it ipso facto} to $\hat t$, while she cannot know which of $\hat T_1$ or $\hat T_2$ is true. %In this way, the common core theory resolves the logical incompatibility of the interpretations of two theory formulations, by the adoption of an internal interpretation that captures their common aspects, and assigns no theoretical or physical significance to the rest.

This does not prevent the scientific realist from, in addition, being committed to one of the two external interpretations, i.e.~committing herself to one of the two theory formulations (perhaps using alternative assessment criteria, i.e.~point (ii) in Section \ref{Qbeudt}): $\hat T_1$, say. Indeed, $\hat t$ and $\hat T_1$ are of course {\it compatible,} by Eq.~\eq{implications}: and so, there is no under-determination here either. 

In any case, the scientific realist does not make a mistake if (perhaps in the absence of additional assessment criteria) she commits only to $\hat t$. % (since, if she commits herself to either $\hat T_1$ or $\hat T_2$, she is {\it ipso facto} also committed to $\hat t$). 
Thus the cautious scientific realist is, in the cases under consideration, {\it always justified in believing the common core theory.}

This is a conclusion that the multiple interpretative options considered in Le Bihan and Read (2018:~Figure 1) obscures. They also consider a ``common core'' theory, but theirs is not constructed by abstraction: in any case, they do not consider the possibility of a constraint, Eqs.~\eq{implications} and \eq{implication0}, that follows from the process of abstraction.\footnote{This is clear from e.g.~the following quote (p.~5): `[T]here is a sense in which, absent further philosophical details, such a move [i.e.~identifying the common core] has made the situation {\it worse}: we have, in effect, identified a {\it further} world which is empirically adequate to the actual world'.} This constraint\footnote{Together with the assumption that {\it at least one of} $\hat T_1$, $\hat T_2$, $\hat t$ is worth scientific realist commitment---else there is no question of under-determination either!} eliminates four out of six of the options in their taxonomy. The remaining two are the two cases just discussed, i.e.~commitment only to $\hat t$, or possible commitment to both $\hat t$ and $\hat T_1$. Thus, as just argued, the problem of under-determination is thereby resolved.\footnote{It is of course not {\it guaranteed} that a common core theory $\hat t$ always exists. However, as I argued in Section \ref{abstr}, this does not lead to cases of empirical under-determination.}

In sum, a scientific realist may lack the resources to know which of two dual theory formulations, $\hat T_1$ or $\hat T_2$, describes the world, but is not {\it required} to believe either of them. Rather, the cautious scientific realist is justified in adopting an internal interpretation, which abstracts from the external interpretations, and thereby resolves their incompatibilities. For unextendable theories, a scientific realist may take theoretical equivalence to be a criterion of theory individuation. Thus a catious scientific realism recommends only belief in the internal interpretation of unextendable theories. \\
%This agrees with earlier discussions of dualities, which have argued that one is justified in being a realist about the internal interpretations common to dual theory formulations of unextendable theories (also in the weak sense), often under additional ontological requirements.
\\
My overall view can be well summarised in terms of how it clarifies previous discussions of under-determination for dualities: by
%Under-determination in the context of dualities was discussed in early work by 
Matsubara (2013) and Rickles (2017). Matsubara presents two approaches to dualities: his first approach (`Accept the different dual descriptions as describing two different situations') corresponds to the Schema's case of theoretically inequivalent theories, and he says that these theoretically inequivalent theories are nevertheless empirically equivalent (`the world may in reality be more like one dual description than the other but we have no empirical way of knowing this'). 

While I broadly agree with Matsubara, my analysis makes several clarifications. For example, my characterisation of the under-determination is slightly different: there is an under-determination specific to dualities only in the case of unextendable theories, but one is always justified in adopting internal interpretations. (For extendable theories, further theory development can distinguish between two theory formulations, so this is ordinary transient under-determination.) Since, in the relevant examples, these are obtained by abstraction from the external interpretations of the various theory formulations, they do not contradict them (Eq.~\eq{implications}). Thus I disagree with Matsubara's conclusion that `This means that we must accept epistemic anti-realism since in this situation it is hard to find any reason for preferring one alternative before another.' Indeed, as I have argued, a cautious scientific realist does not {\it need} to choose one dual over the other: such a realist is justified in adopting an internal interpretation of the kind discussed, which agrees---in everything it says---with the external interpretations, and is silent about their other specific aspects.

Matsubara's second approach (`We do not accept that [the duals] describe different situations; instead they are descriptions of the same underlying reality') corresponds to my case of theoretically equivalent duals. I agree with Matsubara that there is no under-determination, but I think this position is less tangled with ontic structural realism than he seems to think (although it is of course compatible with it). As the notions of an internal interpretation and its justification through unextendability should make clear, judging two duals as being theoretically equivalent is a matter of setting both formal {\it and} interpretative criteria for individuating theories. \\

In the case of unextendable theories, a methodological preference for internal interpretations surfaces---because justified by unextendability and abstraction. This feeds into physicists' main interest in dualities, which comes from their heuristic power in formulating new theories. On an internal interpretation, a duality is thus taken to be a starting point of theory individuation, with an interpretation of the theory's common core, of which the various theory formulations are specific instantiations. This further motivates De Haro and Butterfield's (2017) proposal to lift the usage of `theory' and `model' ``one level up'', so that the various theory formulations are `models' (i.e.~instantiations or representations) of a single theory. And this now holds not only for bare theories and models, but also---on internal interpretations---for interpreted ones.\footnote{My conclusions are similar to some of Dawid's (2017b). Where he says that there is a shift in the role played by empirical equivalence in theory construction (and I agree with this, see De Haro (2018)), I have argued that the traditional semantic and syntactic construals of empirical equivalence are in themselves sufficient to analyse dualities, and that the main difference is in the {\it theories} whose empirical equivalence is being assessed.}

\section{Conclusion}\label{conclusion}

The discussion of under-determination begins slightly differently, and is complementary, to usual discussions of duality. While usual discussions of duality aim to establish conditions for when duals are theoretically equivalent, when analysing under-determination we begin with theoretically {\it inequivalent} theories, whose inequivalence---as my analysis in Section \ref{physin} shows---ends up depending on {\it external} interpretations. Thus our discussion has put two under-emphasised aspects of duality to good use: namely, theoretical inequivalence and external interpretations. %However, the resolution of the under-determination is again in a common core theory, obtained by a process of abstraction.

The Schema's construal of bare theories, interpretation, and duality prompts a notion of theoretical equivalence that, although strict, turns out to generate cases of under-determination: but I have also argued that these do not present a new problem for a catious scientific realism. For theory formulations that can be extended beyond their domain of application, we have the familiar situation of transient under-determination, where further theory development may be expected to break the duality. For theory formulations that cannot be so extended (or only in a way that does not change their interpretations in the old domain), the under-determination is benign, because---for dualities in the literature that are sufficiently well-undestood---a common core theory can be obtained by abstraction from the external interpretations. In this case, a cautious scientific realism does not commit to the external interpretations, but belief in the internal interpretation is justified. Thus the under-determination is here benign, and dualities do not provide the anti-realist with new ammunition, although they may be taken to favour a cautious approach to scientific realism.

Just as dualities bear on the problem of theory individuation that is central to the discussion of theoretical equivalence, they also bear on the question of scientific realism. Namely, they suggest a cautious approach, according to which inter-theoretic relations should be taken into account when determining one's realist commitments.
%be explored before one's realist commitments are fully determined. 

%I also argued, in Section \ref{eed}, that my notion of `empirical equivalence', although liberal (and perhaps surprising), is straightforward: and it is sanctioned by two main notions of empirical equivalence, viz.~van Fraassen's semantic, and Quine's syntactic notions. What we did to achieve this was to apply van Fraassen's isormorphism to duality, and apply Quine's notion to theories with non-standard interpretations. Although the result seems in itself surprising, some such result might have been expected from dualities.

%Most previous studies took dualities to relate theories that are empirically equivalent. %: even in cautious authors who judge that duals are not invariably physically equivalent (this extends to some of the recent literature on theoretical equivalence cited also). 
%As the analysis of Section \ref{eed} should make clear, empirical equivalence is not automatic either. It requires the just-mentioned footwork.

%Surely one of the most surprising and scientifically significant aspect of dualities is that their external interpretations admit translations that make the theories empirically equivalent. I hope that the analysis presented here serves to clarify the notion of empirical equivalence in contemporary physics.

\section*{Acknowledgements}
\addcontentsline{toc}{section}{Acknowledgements}

I thank Jeremy Butterfield, James Read, and two anonymous reviewers for comments on this paper. I also thank John Norton for a discussion of duality and under-determination. This work was supported by the Tarner scholarship in Philosophy of Science and History of Ideas, held at Trinity College, Cambridge.

\section*{References}
\addcontentsline{toc}{section}{References}

\small

Ammon, M.~and Erdmenger, J.~(2015). {\it Gauge/Gravity Duality. Foundations and Applications.} Cambridge: Cambridge University Press.\\
\\
Balasubramanian, V., Kraus, P., Lawrence, A.~E.~and Trivedi, S.~P.~(1999). `Holographic probes of anti-de Sitter space-times'. {\it Physical Review D,} 59, p.~104021.\\
\\
%Barrett, T.~W.~(2018). `Equivalent and Inequivalent Formulations of Classical Mechanics'. PhilSci: http://philsci-archive.pitt.edu/13092.\\
%\\
Barrett, T.~W.~and Halvorson, H.~(2016). `Glymour and Quine on theoretical equivalence'. {\it Journal of Philosophical Logic}, 45 (5), pp.~467-483.\\
\\
Baxter, R.~J.~(1982). {\it Solved Models in Statistical Mechanics}.  Academic Press.\\
\\
Butterfield, J.~(2020). `On Dualities and Equivalences Between Physical Theories'. Forthcoming in {\it Space and Time after Quantum Gravity}, Huggett, N.~and W\"uthrich, C.~(Eds.). An extended version is in: http://philsci-archive.pitt.edu/14736.\\
\\
Castellani, E.~and Rickles, D.~(2017). `Introduction to special issue on dualities'. {\it Studies in History and Philosophy of Modern Physics}, 59: pp.~1-5. doi.org/10.1016/j.shpsb.2016.10.004.\\
\\
Coffey, K.~(2014). `Theoretical Equivalence as Interpretative Equivalence'. {\it The British Journal for the Philosophy of Science,} 65 (4), pp.~821-844.\\
\\
Dawid, R.~(2006). `Under-determination and Theory Succession from the Perspective of String Theory'. {\it Philosophy of Science,} 73 (3), pp.~298-322.\\
\\
Dawid, R.~(2017a). `Scientific Realism and High-Energy Physics'. {\it The Routledge Handbook of Scientific Realism,} J.~Saatsi (Ed.), pp.~279-290.\\
\\
Dawid, R.~(2017b). `String Dualities and Empirical Equivalence'. {\it Studies in History and Philosophy of Modern Physics,} 59, pp.~21-29.\\
\\
De Haro, S.~(2017). `Dualities and Emergent Gravity: Gauge/Gravity Duality'. {\it Studies in History and Philosophy of Modern Physics,} 59, pp.~109-125. arXiv:1501.06162 [physics.hist-ph]\\
\\
De Haro, S.~(2019). `Theoretical Equivalence and Duality'. {\it Synthese,} topical collection on Symmetries. M. Frisch, R. Dardashti, G. Valente (Eds.), 2019, pp. 1-39. arXiv:1906.11144 [physics.hist-ph]\\
\\
De Haro, S. (2020). `Spacetime and Physical Equivalence'. In: {\it Beyond Spacetime. The Foundations of Quantum Gravity,} Huggett, N., Matsubara, K.~and Wüthrich, C.~(Eds.), pp.~257-283. Cambridge: Cambridge University Press. arXiv:1707.06581 [hep-th].\\
\\
De Haro, S.~(2020a). {\it On Inter-Theoretic Relations and Scientific Realism.} PhD dissertation, University of Cambridge. http://philsci-archive.pitt.edu/17347.\\
\\
De Haro, S.~(2020b). `On Empirical Equivalence and Duality'. To appear in {\it 100 Years of Gauge Theory. Past, Present and Future Perspectives,} S.~De Bianchi and C.~Kiefer (Eds.). Springer. arXiv:2004.06045 [physics.hist-ph].\\
\\
De Haro, S.~and Butterfield, J.N.~(2017). `A Schema for Duality, Illustrated by Bosonization'. In: Kouneiher, J.~(Ed.), {\it Foundations of Mathematics and Physics one century after Hilbert}, pp.~305-376. arXiv:1707.06681 [physics.hist-ph].\\
%%CITATION = ARXIV:1707.06681;%%
\\
De Haro, S., Mayerson, D.~R.~and Butterfield, J.~(2016). `Conceptual Aspects of Gauge/Gravity Duality'. {\it Foundations of Physics,} 46, pp.~1381-1425. arXiv:1509.09231 [physics.hist-ph].\\
\\
Dieks, D., Dongen, J. van, Haro, S. de~(2015). `Emergence in Holographic Scenarios for Gravity'. 
%PhilSci 11271, arXiv:1501.04278 [hep-th]. 
{\it Studies in History and Philosophy of Modern Physics} 52 (B), pp.~203-216. arXiv:1501.04278 [hep-th].\\ %doi:~10.1016/j.shpsb.2015.07.007.\\
  %%CITATION = ARXIV:1501.04278;%% 
%\\
\\
Earman, J.~(1993). `Underdetermination, Realism, and Reason'. {\it Midwest Studies in Philosophy,} XVIII, pp.~19-38.\\
\\
Festuccia, G.~and Liu, H.~(2006). `Excursions beyond the horizon: Black hole singularities in Yang-Mills theories. I', {\it Journal of High-Energy Physics,} 04, 044.\\
\\
Glymour, C.~(1970). `Theoretical Equivalence and Theoretical Realism. PSA: Proceedings of the Biennial Meeting of the Philosophy of Science Association 1970, pp.~275-288.\\
\\
Glymour, C.~(1977). `The epistemology of geometry'. {\it No$\hat u$s}, pp.~227-251.\\
\\
Hodges, W.~(1997). {\it A Shorter Model Theory.} Cambridge: Cambridge University Press.\\
\\
%Hudetz, L.~(2019). `Definable Categorical Equivalence'. PhilSci: http://philsci-archive.pitt.edu/14297.\\
%\\
Huggett, N.~(2017). `Target space $\neq$ space'. {\em Studies in History and Philosophy of Modern Physics}, 59, 81-88.\\ % doi:10.1016/j.shpsb.2015.08.007.\\
\\
Huggett, N.~and W\"uthrich, C.~(2020). {\it Out of Nowhere: Duality,} Chapter 7. Oxford: Oxford University Press (forthcoming).
http://philsci-archive.pitt.edu/17217.\\
\\
Kitcher, P.~(1993). {\it The Advancement of Science.} New York and Oxford: Oxford University Press.\\
\\
Laudan, L.~(1990). `Demystifying Underdetermination'. {\it Minnesota Studies in the Philosophy of Science,} 14, pp.~267-297.\\
\\
Laudan, L.~and Leplin, J.~(1991). `Empirical Equivalence and Underdetermination'. {\it The Journal of Philosophy,} 88 (9), pp.~449-472.\\
\\
Le Bihan, B.~and Read, J.~(2018). `Duality and Ontology'. {\it Philosophy Compass,} 13 (12), e12555.\\
\\
Lenzen, V.~F.~(1955). `Procedures of Empirical Science'. In: {\it International Encyclopedia of Unified Science,} Neurath, O., Bohr, N., Dewey, J., Russell, B., Carnap, R., and Morris, C.~W.~(Eds.). Volume I, pp.~280-339.\\
\\
%Lutz, S.~(2017). `What Was the Syntax-Semantics Debate in the Philosophy of Science About?' {\it Philosophy and Phenomenological Research}, XCV (2), pp.~319-352.\\
%\\
Lyre, H.~(2011). `Is Structural Underdetermination Possible?' {\it Synthese,} 180, pp.~235-247.\\
\\
Matsubara, K.~(2013). `Realism, Underdetermination and String Theory Dualities'. {\it Synthese,} 190, 471-489.\\
\\
Ney, A.~(2012). `Neo-Positivist Metaphysics'. {\it Philosophical Studies,} 160, pp.~53-78.\\
\\
Norton, J.~(2006). `How the Formal Equivalence of Grue and Green Defeats What Is New in the New Riddle of Induction'. {\it Synthese}, 150, pp.~185-207.\\
\\
%Norton, J.~(2008). `Must Evidence Underdetermine Theory?' In: {\it The Challenge of the Social and the Pressure of Practice: Science and Values,}, M.~Carrier, D.~Howard, J.~A.~Kourany (Eds.). University of Pittsburgh Press.\\
%\\
Polchinski, J.~(2005). {\it String Theory,} volume II. Cambridge: Cambridge University Press. Second Edition.\\
\\
Psillos, S.~(1999). {\it Scientific Realism. How Science Tracks Truth.} London and New York: Routledge.\\
\\
%Quine, W.~V.~(1960). {\it Word and Object,} New Edition. Cambridge, MA and London: The MIT Press.\\
%\\
Quine, W.~V.~(1970). `On the Reasons for Indeterminacy of Translation'. {\it The Journal of Philosophy,} 67 (6), pp.~178-183.\\
\\
Quine, W.~V.~(1975). `On empirically equivalent systems of the world'. {\it Erkenntnis}, 9 (3), pp.~313-328.\\
\\
Read, J.~(2016). `The Interpretation of String-Theoretic Dualities'. {\it Foundations of Physics,} 46, pp.~209-235.\\
\\
Read, J.~and M\o ller-Nielsen, T.~(2020). `Motivating Dualities'. {\it Synthese,} 197, pp.~263-291.\\
\\
Rickles, D.~(2011). `A Philosopher Looks at String Dualities'. {\it Studies in History and Philosophy of Modern Physics,} 42, pp.~54-67.\\
\\
Rickles, D.~(2017). `Dual Theories: `Same but Different' or `Different but Same'?' {\it Studies in History and Philosophy of Modern Physics,} 59, pp.~62-67. \\
\\
Schwarz, J.~(1992). `Superstring Compactification and Target Space Duality'. In: N.~Berkovits, H.~Itoyama, K.~Schoutens, A.~Sevrin, W.~Siegel, P.~van Nieuwenhuizen and J.~Yamron, {\it Strings and Symmetries.} World Scientific, pp.~3-18. \\
%%CITATION = HEP-TH/9108022;%%
\\
Sklar, L.~(1975). `Methodological Conservatism'. {\it The Philosophical Review,} 84 (3), pp.~374-400.\\
\\
Stanford, P.~K.~(2006). {\it Exceeding Our Grasp.} New York: Oxford University Press.\\
\\
van Fraassen, B.~C.~(1970). `On the Extension of Beth's Semantics of Physical Theories'. {\it Philosophy of Science,} 37 (3), pp.~325-339.\\
\\
van Fraassen, B.~C.~(1980). {\it The Scientific Image}. Oxford: Clarendon Press.\\
\\
%van Fraassen, B.~C.~(1989). {\it Laws and Symmetry}. Oxford: Clarendon Press.\\
%\\
%Weatherall, J.~O.~(2015). `Categories and the Foundations of Classical Field Theories'. PhilSci: http://philsci-archive.pitt.edu/11587.\\
%\\
Weatherall, J.~O.~(2020). `Equivalence and Duality in Electromagnetism'. Forthcoming in {\it Philosophy of Science.} https://doi.org/10.1086/710630.\\
\\
Zwiebach, B.~(2009). {\it A First Course in String Theory.} Cambridge: Cambridge University Press.

\end{document}